\documentclass[sigconf]{acmart}
\usepackage[british]{babel}

\def\BibTeX{{\rm B\kern-.05em{\sc i\kern-.025em b}\kern-.08emT\kern-.1667em\lower.7ex\hbox{E}\kern-.125emX}}
    
\copyrightyear{2019} 
\acmYear{2019} 
\setcopyright{acmlicensed}
\acmConference[ARES '19]{Proceedings of the 14th International Conference on Availability, Reliability and Security (ARES 2019)}{August 26--29, 2019}{Canterbury, United Kingdom}
\acmBooktitle{Proceedings of the 14th International Conference on Availability, Reliability and Security (ARES '19), August 26--29, 2019, Canterbury, United Kingdom}
\acmPrice{15.00}
\acmDOI{10.1145/3339252.3340334}
\acmISBN{978-1-4503-7164-3/19/08}

\usepackage{amsmath}
\usepackage{url}
\usepackage{graphicx}
\usepackage{gensymb}
\usepackage{makecell}
\usepackage{todonotes}
\graphicspath{{../png/}{../pdf/}{../jpeg/}}
\DeclareGraphicsExtensions{.pdf,.jpeg,.png}
\usepackage{array}

\usepackage[flushleft]{threeparttable}

\begin{document}

%
\title[Privacy-Enhancing Context Authentication from Location-Sensitive Data]{Privacy-Enhancing Context Authentication from Location-Sensitive Data}

%
\author{Pradip Mainali}
\email{pradip.mainali@onespan.com}
\affiliation{%
  \institution{OneSpan}
  \city{Brussels}
  \country{Belgium}}

\author{Carlton Shepherd}
\email{carlton.shepherd@onespan.com}
\affiliation{%
  \institution{OneSpan}
  \city{Cambridge}
  \country{United Kingdom}}

\author{Fabien A. P. Petitcolas}
\email{fabien.petitcolas@onespan.com}
\affiliation{%
  \institution{OneSpan}
  \city{Brussels}
  \country{Belgium}}

%
\renewcommand{\shortauthors}{P. Mainali, C. Shepherd, and F. A. P. Petitcolas}

%
\begin{abstract}

This paper proposes a new privacy-enhancing, context-aware user authentication system, ConSec, which uses a transformation of general location-sensitive data, such as GPS location, barometric altitude and noise levels, collected from the user's device, into a representation based on locality-sensitive hashing (LSH). The resulting hashes provide a dimensionality reduction of the underlying data, which we leverage to model users' behaviour for authentication using machine learning.  We present how ConSec supports learning from categorical \emph{and} numerical data, while addressing a number of on-device and network-based threats. 
ConSec is implemented subsequently for the Android platform and evaluated using data collected from 35 users, which is followed by a security and privacy analysis. We demonstrate that LSH presents a useful approach for context authentication from location-sensitive data without directly utilising plain measurements.
\end{abstract}

%
%


%

%
\maketitle

\section{Introduction}
Secure and usable mobile authentication mechanisms are critical to numerous applications, including internet banking, email, and social media. Traditional knowledge-based authentication mechanisms, such as PINs, patterns and passwords, remain fraught with concerns surrounding memorability, credential reuse, sufficient entropy, and shoulder-surfing attacks~\cite{pearman2017let,stajano2011pico}. Meanwhile, token-based methods, e.g. one-time password (OTP) generators and key fobs, can be lost or stolen and are generally costly to produce, maintain and replace~\cite{o2003comparing}.
Biometrics, such as fingerprint and facial authentication, have become widely-deployed on modern mobile handsets; however, environmental factors, like moisture, injury, skin complexion, and lighting can significantly increase error rates~\cite{de2015secure}. Traditional approaches also feature an overarching drawback: the user is authenticated only once, after which all access to assets and services and granted thereafter.  Such all-or-nothing authentication has prompted both security and usability concerns in related literature~\cite{hayashi_casa:_2013,riva2012progressive,harbach2014sa,gupta2012intuitive}. 

Context authentication systems have arisen from these concerns, which transparently compute an authentication score from device data, such as GPS location, nearby Wi-Fi access points (APs), and cellular network information, with respect to previously observed behaviour~\cite{shi_senguard:_2011,hayashi_casa:_2013,gupta2012intuitive,conXsense}. Such systems may infer the user's authentication status directly,  i.e. accept/reject~\cite{shi_senguard:_2011}, or determine access control policies and explicit authentication strength~\cite{hayashi_casa:_2013,conXsense}. Unfortunately, many context authentication systems necessitate collecting swathes of privacy-sensitive data. This is especially problematic if the scheme is deployed by a remote authentication service and the data is later disclosed in a security breach or misused without consent. Several surveys have already demonstrated users' reticence in disclosing location data used by many context authentication proposals~\cite{fisher_short_2012-1,danezis_how_2005,barkhuus_location-based_2003}.  
Fisher-Short et al.~\cite{fisher_short_2012-1} show, for instance, that users tend to deny mobile applications the permission to access location data because of concerns about it being transmitted remotely and shared surreptitiously with third-parties.

In light of this discussion, we present a novel context authentication system, ConSec, which enhances the confidentiality of device data using the Super-Bit Locality-Sensitive Hashing (SB-LSH) proposal by Ji et al.~\cite{ji_super-bit_2012}. 
SB-LSH is a dimensionality reduction algorithm wherein the locality-sensitive hashes reveal only the relative distance between user location measurements without disclosing their precise location on Earth; from this, machine learning can be applied to model the user's behaviour.  In this work, we demonstrate how ConSec enables computation over numerical and categorical data, e.g. Wi-Fi ESSIDs, to flexibly support various modalities for privacy-enhancing contextual authentication.  We evaluate ConSec using data collected from 35 participants, after which we analyse its effectiveness against certain attacks, such as triangulation, and their effect on error rates.   

The main contributions of this paper is a new approach to contextual authentication by protecting the confidentiality of numerical location-sensitive data, such as GPS coordinates, using SB-LSH~\cite{ji_super-bit_2012}. ConSec can learn user authentication models from protected numerical and categorical data using standard machine learning algorithms with low error rates.
This paper begins with a review of existing contextual and privacy-enhancing continuous authentication schemes in Section~\ref{sec:relatedwork}.  Section~\ref{sec:consecarchitecture} then describes the architecture of the ConSec, including the threat model and the modalities used.  Sections~\ref{sec:enhancing_privacy_contextual_data} and~\ref{sec:contextauthenticationalgorithm} describe the use of LSH and keyed HMACs for privacy-preserving behavioural modelling from numerical and categorical data respectively for contextual authentication.   Next, an evaluation of the proposed scheme is presented from a user base of 35 users in Section~\ref{sec:experimentalresults}, which is followed by a security and privacy analysis in Section~\ref{sec:privacyanalysis}.  Lastly, Section~\ref{sec:conclusion} concludes this paper, including a discussion of future research directions.


\section{Related Work}\label{sec:relatedwork}
This section explores the state-of-the-art of \emph{contextual authentication}, focused on in this work, and \emph{privacy-enhancing continuous authentication} generally. We summarise notable proposals and their contributions.

\subsection{Contextual Authentication}
\label{rel:contextual_authentication}

The CASA system by Hayashi et al.~\cite{hayashi_casa:_2013} focuses on reducing the strength of user-authentication challenges based on the current location of the user and previously observed behaviour. The authentication strength is determined by the presence of the user in location of varying perceived risk, such as at home or at work. CASA applies a Na\"{i}ve Bayes classifier to model user behaviour from GPS location, which produces a probability value based on their past behaviour to determine the device's explicit authentication strength, e.g. password, PIN or even no challenge.  The system is trialled using 32 users, with a location classification accuracy of 92\%, and 68\% of explicit authentication attempts being reduced.


SenGuard by Shi et al.~\cite{shi_senguard:_2011} uses five modalities---location, cell tower ID, voice, touch, and motion-based activity recognition (cycling, walking, stationary, etc.)---from which a multitude of features are extracted, including touch gestures, GPS correlation, and Levenshtein distance of cell IDs in a sliding window. From this, the system computes an aggregated authentication decision from individual, modality-specific classifiers, which is used to authenticate the user in a binary fashion, i.e. accept or reject.  SenGuard is evaluated with a user base of four participants, yielding a classification accuracy of 95.8--97.1\% depending on the user.

Gupta et al.~\cite{gupta2012intuitive} present the first proposal for setting mobile device access control policies based on users' contexts. The authors use features based on Context of Interest (CoI), corresponding to locations a user visits frequently or spends significant amounts of time in.  CoIs are generated from clusters of GPS coordinates, Wi-Fi access points (APs) and nearby Bluetooth devices, which are compared with past data using similarity metrics, such as the set intersection, using manually set threshold values.  The CoIs are then divided into `safe' and `unsafe' locations, such as `at home' and `out shopping' respectively, which are used for dynamically setting access control policy profiles.  An evaluation with 37 users resulted in average precisions of 0.854 (safe locations) and 0.311 (unsafe), and recalls of 0.917 (safe) and 0.341 (unsafe).

Miettenen et al.~\cite{conXsense} present ConXsense, which uses contextual data for setting risk-based device locks, akin to~\cite{hayashi_casa:_2013}, as well as access control policy profiles.  The authors also generate CoI features, but, unlike \cite{gupta2012intuitive}, these are inputted to a classifier with ground truth labels drawn from user input.  The authors evaluate the system using data from 15 users and Random Forest, k-Nearest Neighbour (kNN), and Na\"{i}ve Bayes classifiers, with an approximate 0.70 true positive rate (TPR) and 0.10 false positive rate (FPR).


Witte et al.~\cite{witte_context-aware_2013} propose a system for scoring authentication behaviour based on the device's GPS location, accelerometer, magnetic field, light, battery and sound measurements. Statistical features are extracted from the raw measurements, such as the arithmetic mean, median, maximum and minimum values, which are aggregated with system-level features, such as screen status and boot and shutdown times. Feature vectors are inputted to an SVM classifier trained on user data collected over a three-day period; an evaluation is performed using data from 15 participants, with an average F1-score of 0.85.

\subsection{Privacy-Enhancing Continuous Authentication}
\label{rel:privacy_enhancing_continuous_authentication}

Shahandashti et al.~\cite{shahandashti2015reconciling} propose a scheme for computing authentication decisions remotely on an honest-but-curious server from encrypted feature vectors using Paillier homomorphic encryption~\cite{Paillier_1999}.  Encrypted behavioural models are stored on a remote server and a decision is computed homomorphically from encrypted feature vectors transmitted from the user's device without exposing the plaintext measurements.  The decision is computed from the dissimilarity between incoming features and the stored user profile using the average absolute distance (AAD); the user is authenticated if the AAD similarity falls within a threshold determined by the service provider. However, while a security proof is provided, the scheme is neither implemented nor evaluated in practice.  

Domingo-Ferrer et al.~\cite{domingo2015flexible} develop the work from~\cite{shahandashti2015reconciling} by presenting an additional approach based on the set intersection of homomorphically encrypted feature vectors to authenticate users. The set intersection is used to compute the dissimilarity function between the encrypted user profile and incoming feature vectors to support categorical features, beyond only numerical features supported in \cite{shahandashti2015reconciling}.  The authors provide a performance evaluation in which authentication takes 0.08--31.2 seconds depending on the number of input features (1--50 features respectively). 

Sedenka et al.~\cite{sedenka2015secure} construct protocols for the outsourcing of the scaled Manhattan (L1) and Euclidean (L2) distances and principal component analysis (PCA) using garbled circuits and homomorphic encryption for use in continuous authentication.  The work tackles the case of computing the similarity of incoming feature vectors with stored user models in the presence of an honest-but-curious server. The authors provide a security analysis and performance evaluation using a consumer laptop and smartphone, the results of which yield a communication and time penalty of 4--174MB and 0.85s--45.9s respectively based on the submitted feature vector size.

Halunen and Vallivaara~\cite{halunen2016secure} present a privacy-enhancing continuous authentication system based on keystroke dynamics with order-preserving symmetric encryption (OPSE) and, like~\cite{shahandashti2015reconciling}, Paillier homomorphic encryption.  The proposal extracts four features regarding the down-down, up-down and down-up times of each keystroke, along with the entered string.  Next, the AAD is computed homomorphically between the sample and the stored template values, while OPSE is used for comparing samples with various thresholds to fine-tune security and usability.  An evaluation is conducted with 20 users yielding a best-case accuracy of 91.5\%.

\subsection{Discussion}

Many existing contextual authentication proposals, summarised in Table~\ref{tab:related}, use device measurements in unprotected form---giving rise to privacy concerns especially when authentication decisions are outsourced to a remote server~\cite{shahandashti2015reconciling,domingo2015flexible,sedenka2015secure}.  As noted previously, studies have already shown that users are generally reluctant towards disclosing, in particular, their GPS location to device applications~\cite{fisher_short_2012-1,danezis_how_2005,barkhuus_location-based_2003,cvrcek_study_2006}.  Some users also disable GPS depending on the sensitivity of the location they visit, irrespective of the application~\cite{barkhuus_location-based_2003}. Work in \cite{fisher_short_2012-1} also shows that, when permission is denied to an application that accesses GPS location, it is principally from concerns relating to whether location data is stored securely (on-device and remotely) and whether it is shared with other parties without consent.

\begin{table}
\caption{Related contextual authentication schemes.}
\begin{threeparttable}
\begin{tabular}{@{}lcc@{}}
\toprule
\textbf{Proposals} & \textbf{Data Modalities} & \textbf{Error Rates} \\\midrule
CASA~\cite{hayashi_casa:_2013}  & GPS Location &    92\% Acc.      \\[0.15cm] 
SenGuard~\cite{shi_senguard:_2011} & \makecell{GPS Location, Voice,\\Touch, AR} &     95.8--97.1\% Acc.        \\[0.3cm]
Gupta et al.~\cite{gupta2012intuitive} & \makecell{GPS Location, Wi-Fi APs,\\Bluetooth Devices} & \makecell{0.311--0.854 Pr.,\\ 0.341--0.917 Re.}           \\[0.3cm]
ConXsense~\cite{conXsense} & \makecell{GPS Location, Wi-Fi APs,\\Bluetooth Devices} & $\sim$70\% TPR, 10\% FPR     \\[0.3cm] 
Witte et al.~\cite{witte_context-aware_2013} & \makecell{GPS Location, Sys., AC,\\MF, Sound, Light} & $\sim$0.85 F1-score \\
\bottomrule
\end{tabular}
\begin{tablenotes}
\item AC: Accelerometer, MF: Magnetic Field, Sys.: System Data, Acc.: Classification Accuracy, FPR: False Positive Rate, TPR: True Positive Rate, AR: Activity Recognition, Pr.: Precision, Re.: Recall.
\end{tablenotes}
\end{threeparttable}
\label{tab:related}
\end{table}



Current approaches to privacy-enhancing continuous authentication have employed homomorphic encryption, e.g. Paillier~\cite{Paillier_1999}, and garbled circuits for two-party computation (2PC)~\cite{halunen2016secure,shahandashti2015reconciling,domingo2015flexible,sedenka2015secure}.  However, challenges still remain with respect computational complexity and storage overhead, which is exemplified by worst cases of 31.2s and 45.9s to compute authentication decisions in~\cite{domingo2015flexible} and~\cite{sedenka2015secure} respectively.  Storage complexity is also an issue, with 4--175MB required for computing a single authentication decision in~\cite{sedenka2015secure}. Homomorphic encryption-based schemes also face inherent challenges regarding the supported arithmetic operations for learning from data, i.e. \emph{only} additions for Paillier-based proposals. Fully homomorphic encryption (FHE)~\cite{gentry2009fully}, meanwhile, has been deemed too cumbersome in the literature for high-frequency, high-dimensional classification tasks like continuous authentication~\cite{sedenka2015secure,shahandashti2015reconciling}.

Lastly, we note that the work in~\cite{halunen2016secure} focuses on keystroke-based authentication, which necessitates user interaction.  Rather, the focus of this work is context authentication from device sensors \emph{without} interaction from the user. 
Ultimately, we aim to demonstrate a simpler approach to privacy-enhancing, zero-interaction context authentication based on keyed HMACs and the Super-Bit Locality-Sensitive Hashing (SB-LASH) algorithm by Ji et al.~\cite{ji_super-bit_2012}, without resorting to the complexities of homomorphic encryption and multi-party computation used in existing solutions.

\section{System Architecture} \label{sec:consecarchitecture}
In this section, we present the system architecture of ConSec for performing context authentication, beginning with a high-level overview before describing the threat model and data modalities considered in this work.

\subsection{System Overview}

\begin{figure*}
	\centering
	\includegraphics[width=0.9\linewidth, interpolate=true]{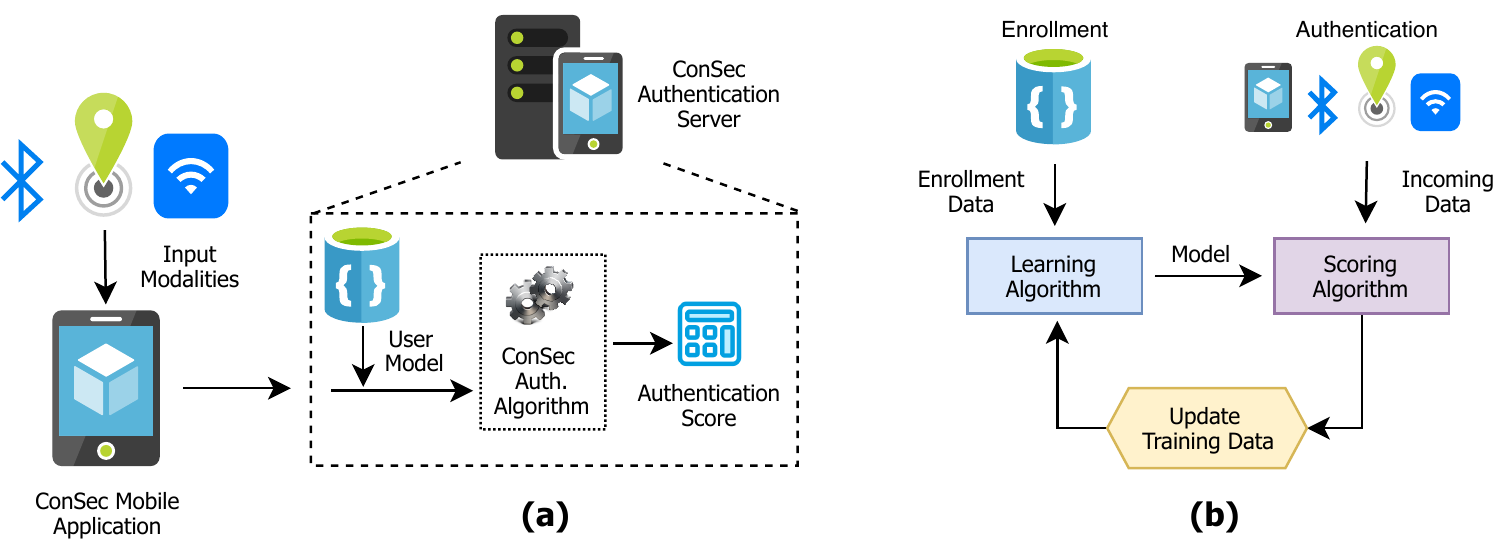}
	\caption{High-level architecture: (a) ConSec context-aware authentication flow; (b) ConSec authentication algorithm.}
	\label{fig:contextual_authentication_blockdiagram}
\end{figure*}

Figure~\ref{fig:contextual_authentication_blockdiagram} illustrates the block diagram of ConSec, comprising the ConSec-App on the smartphone for data collection and transmission, and the authentication algorithm (ConSec-Auth) that executes remotely on the authentication server. ConSec-App transforms \underline{\emph{categorical}} contextual data, such as nearby Wi-Fi ESSIDs and MAC addresses, using HMAC-SHA256 under a randomly generated, per-user key within the application, which we denote $HK$.  \underline{\emph{Numerical}} location-sensitive data, such as the device's GPS location and barometric altitude, is transformed using the Super-Bit LSH (SB-LSH) algorithm proposed in~\cite{ji_super-bit_2012}. The transformed HMAC and SB-LSH values are transmitted to ConSec-Auth, which computes the authentication score based on the user's previously observed behaviour. We return to the complete set of contextual data collected by ConSec-App in Section~\ref{sec:contextual_data}.   

Figure~\ref{fig:contextual_authentication_blockdiagram}(b) shows the block diagram of the ConSec-Auth algorithm, comprising enrollment and authentication phases. We build on the assumption that contextual data generally form clusters around locations frequently visited by the user, such as home, work, and so on, according to their day-to-day mobility patterns.  The enrollment phase constructs the initial model that represents these mobility patterns, which is employed in the authentication phase to compute a real-valued authentication score of the user's newly observed behaviour, $\mathcal{S} \in [0, 1]$, to be used by the service provider.




%
%
%
%
%
%
\subsection{Assumptions and Threat Model}
\label{sec:threat_model}

ConSec aims to provide lightweight, server-side context authentication for informing secure access to remote assets from mobile sensor data. The scheme is intended to allow a service provider to tailor access to sensitive assets based on the output probability without incurring the complexities of existing techniques, such as homomorphic encryption \cite{shahandashti2015reconciling,domingo2015flexible,sedenka2015secure,halunen2016secure}.

Server-side, we assume a remote service operating under the honest-but-curious model, which executes the scheme as intended but attempts to infer users' behaviour based on the device data samples that it observes.  The service may use any insights for unintended purposes, such as profiled advertising or selling the information to unauthorised third-parties without consent. That is, the goal is to provide context authentication for informing server-side access control while precluding the service's ability to easily recover the actual measurements it receives.  However, we note that, even under model whereby the server is wholly honest, the use of privacy-preserving context authentication also limits the impact of security breaches that lead to the unauthorised disclosure of plain-text measurements, and the publicity or legal repercussions that may follow. We also assume that the ConSec mobile application is available in a trusted application store.

On the device, we assume a trusted keystore and the ability to collect, transform and transmit sensor values securely. The keystore is used to hold certificates used to mutually authenticate the end-points for transmitting feature data across a standard network channel, e.g. Wi-Fi, securely using TLS. Both components may be instantiated using conventional security controls provided by modern mobile operating systems, such as application sandboxing, which we concentrate on this work.  However, for further security assurances against kernel-mode (ring 0) adversaries, this may be realised using a trusted execution environment (TEE), as suggested in \cite{liu2012software} and \cite{shepherd2017towards}. We also consider a context-manipulating adversary, e.g. \cite{shrestha2015contextual}, with the ability to influence the measurements collected by the device; for example, by instantiating rogue Wi-Fi APs with spoofed ESSIDs and MAC addresses to maliciously influence the authentication scoring algorithm. This forms the basis of the evaluation described later in Section \ref{sec:experimentalresults}.

\subsection{Contextual Data}\label{sec:contextual_data} 
Modern mobile devices contain a varieties of sensors, such as GPS chips, magnetometers, accelerometers and pressure sensors, and related modules, e.g. Wi-Fi and Bluetooth, that can be used for collecting contextual data. 
The data types used by ConSec are described in the following subsections.


\subsubsection{Geographic Location}
\label{sec:location}
The geographic location of the user is collected from the device's GPS module or, if it is unavailable, using Android's network location, which returns recently observed GPS coordinates, or coordinates based on the Cell ID or Wi-Fi network location in that order~\cite{android:location}.
The geodetic latitude ($\phi$) and longitude ($\lambda$) values are first transformed to the Earth-Centre-Earth-Fixed (ECEF) Cartesian coordinate system accounting for the Earth's ellipsoidal shape using the WGS84 model~\cite{departmentofdefensewgs84_2000}. Equation~\ref{eq:ecef_xyz} is used to transform the geodetic latitude $\phi$, longitude $\lambda$ and altitude $h$ cordinates to the ECEF coordinate system. The coordinates of a point ${\bf{p}}$ in Cartesian coordinates on Earth's surface are given by: 
\begin{equation}
\label{eq:ecef_xyz}
\left(\begin{array}{c}{p}_{x}\\ {p}_{y}\\ {p}_{z}\end{array}\right)=\left(\begin{array}{c}\left(N(\phi) + h\right)\cos(\phi)\cos(\lambda)\\ \left(N(\phi) + h\right)\cos(\phi)\sin(\lambda) \\ \left((b^2/a^2) N(\phi) + h\right) \sin(\phi) \end{array}\right)
\end{equation}
where
\begin{equation}
\label{eq:latitude_s}
N(\phi) = \frac{a}{\sqrt{(1 - e^2\sin^2(\phi))}}
\end{equation}
and the squared first-eccentricity ($e^2$), the semi-major axis ($a$) and the semi-minor axis ($b$) are taken from WGS84: $e^2=6.69437999014\times10^{-3}$, $a=6378137$m and $b=6356752.3142$m.

\subsubsection{Barometric Altitude}

Barometric pressure provides information regarding the height of the device from the sea level as atmospheric pressure is proportional to altitude; it is also measurable within enclosed spaces, e.g. buildings, where GPS coordinates may not be ascertained. 
The barometric altitude is calculated from the measured pressure, $p$, reference pressure, $p_0$, and temperature, $T_0$, using the following formula from~\cite{wallace1977atmospheric}:
\begin{equation}
\label{eq:baro_altitude}
H = \frac{273.15 + T_0}{0.0065}\left(1 - \left(\frac{p}{p_0}\right)^\frac{1}{5.255}\right)
\end{equation}
ConSec-App uses METAR\footnote{METAR: \url{https://www.aviationweather.gov/dataserver}} (Meteorological Aerodrome Report) for acquiring $p_0$ and $T_0$.

\subsubsection{Noise Level}

The average amplitude of the background noise is estimated using a three-second recording from the device's microphone. This provides additional contextual information regarding the user's location, which is also used in~\cite{witte_context-aware_2013} as an authentication feature; for intuition, the device will likely measure a lower amplitude in a quiet office versus a loud city-centre environment.


\subsubsection{Magnetic Fingerprint}
This comprises the geomagnetic field strength and the geomagnetic inclination angle computed from the device's magnetometer and accelerometer sensors. Magnetic field data varies on the Earth's surface---it is strongest at the poles and weakest at the equator---and thus provides some information about the user's location. The magnetic inclination angle is the angle at which the magnetic field lines intersects with the surface of the Earth; it ranges from zero degrees at the equator to 90~degrees at the poles.
The accelerometer data is used to find the orientation of the device with respect to the world coordinate system, and the field data read by the device's magnetometer is transformed to align with respect to the world coordinate system from which the magnetic inclination angle is computed.

\subsubsection{Wi-Fi}
The data collected from Wi-Fi networks includes the name and the MAC address of the router to which the phone is connected, the primary and secondary DNS settings of the router, the received signal strength indication (RSSI), and the list of Wi-Fi names and MAC addresses of nearby access points (APs). 

\subsubsection{Mobile Network} 
The data collected from a mobile network comprises the mobile network type---2G, 3G, and so on---RSSI, operator name, mobile network code (MNC), mobile network country code (MCC), location area code (LAC), and mobile network cell ID. 

\subsubsection{Bluetooth}
The Bluetooth module of ConSec acquires a list of the (visible) Bluetooth device names and MAC addresses detected within the device's vicinity.

\section{Enhancing the Privacy of Contextual Data}\label{sec:enhancing_privacy_contextual_data}
Evidently, location-sensitive data carries private information regarding the user's whereabouts that could be exploited, particularly when authentication decisions are computed remotely. This is a major shortcoming of many existing contextual authentication schemes identified in Section~\ref{sec:relatedwork}.
As such, to enhance the privacy of the location-sensitive information listed previously, \emph{numerical} contextual data is subject to the SB-LSH algorithm by Ji et al.~\cite{ji_super-bit_2012}, while keyed HMACs are used for \emph{categorical} data.  The following sections describe these procedures in further detail.

\subsection{Numerical Data}
\label{lsh:described}
The SB-LSH algorithm~\cite{ji_super-bit_2012} is computed from seven features: the geographic location $\textbf{p} = (p_x, p_y, p_z)$ from Section \ref{sec:location}; the barometric altitude; the noise level; the magnetic field strength; and the magnetic inclination angle. This information is salted with a long-lived, randomly generated, per-user value, $s$, which is used to create an eight-dimensional vector, $\bf{x}$. 

SB-LSH is initialised by generating $K$ eight-dimensional vectors $[\bf{v}_1, \bf{v}_2, \dots, \bf{v}_K]$ sampled from the normal distribution $\mathcal{N}(0, 1)$ and orthogonalised in blocks of eight vectors using the Gram-Schmidt process. The vector, $\bf{x}$, is then projected to produce $h_{\bf{v}_{i}}({\bf{x}})=\mathrm{sign}({\bf{v_{i}}}^T{\bf{x}})$, where $\mathrm{sign}(.)$ is defined as:
\begin{align}
\label{eq:sblsh_hash_sgn}
\mathrm{sign}(z) = & 1, z\ge 0 \nonumber \\
& 0, z<0	
\end{align}
This results in a $K$-bit hash code $H({\bf{x}}) = \left\{h_{\bf{v}_1}({\bf{x}}),h_{\bf{v}_2}({\bf{x}}),\dots,h_{\bf{v}_K}({\bf{x}})\right\}$. It is also shown in~\cite{ji_super-bit_2012} that the Hamming distance between two hash codes, $a$ and $b$, is related to the angular distance, $\theta_{a,b}$:
\begin{equation}
\label{eq:sblsh_hash_hamming}
E\left[d_{\mbox{Hamming}}\left(h(a), h(b)\right)\right] = \frac{K\theta_{a,b}}{\pi}=C\theta_{a,b}
\end{equation}
Where $C=K/\pi$ is constant. As such, the similarity (cosine distance) between the hash values can be computed.  The $K$ randomly initialized LSH vectors are used to compute the location-hash. An issue then arises pertaining to setting the value of $K$, which depends on the security strength and the accuracy that is needed in computing the distance from the location-hash. In both cases, a large value of $K$ is desired; however, on the other hand, a large $K$ increases the amount of data needed to be transmitted and stored on the server. A value of $K$ that gives a reasonable error in computing the distance without compromising security is hence needed.  

\begin{table}
	\centering
	\caption{Error variation for $K$ bits in LSH.}
	\label{tab:location_hash_accuracy}
	\begin{tabular}[t]{ c c c} \toprule
		\textbf{$K$ bits} & \textbf{MAE} & \textbf{RMSE} \\ \midrule
		128 & 41.908 & 65.709 \\ 
		256 & 35.132 & 44.051 \\ 
	    512 & 26.654 & 32.236 \\ 
		1024 & 17.655 & 23.181 \\ 
		2048 & 13.355 & 17.049 \\ 
		4096 & 9.785 & 12.648 \\ 
		8192 & 7.568 & 9.722 \\ 
		16384 & 6.160 & 7.874 \\ \bottomrule
	\end{tabular}
\end{table}

We evaluated this for different values of $K$ as follows. One thousand different location pairs were generated randomly, where the first location value in the pair was generated randomly and the second was generated to be $25$km away from the first. $25$km was chosen as a preliminary value as we assume that general user mobility patterns are limited geographically to small regions, say, for work and home. The true and approximate central angles between the locations in the pair were computed from the actual location values and the location hash codes respectively. The experiment was repeated for ten thousand different SB-LSH hash codes. 
Table~\ref{tab:location_hash_accuracy} shows the mean absolute error (MAE) and root-mean-square error (RMSE) for different $K$ values, averaged over all hash code instances.   The error in computing the distance between the locations can be computed as the central angle between the locations is known, which is computed using the great circle distance formula as follows:
\begin{equation}
\label{eq:distance_from_angle}
d=2\times R\times\sin\left(\frac{\theta}{2}\right)
\end{equation}
Where $\theta$ is the central angle between the actual locations in radians and $R=6371$ km is the mean radius of Earth. The error in computing the distance from location-hashes is given by: 
\begin{equation}
\label{eq:error_in_distance}
d_e=2\times R\times \left(\sin\left(\frac{\theta_t}{2}\right) - \sin\left(\frac{\theta_a}{2}\right)\right)
\end{equation}
Where $\theta_a$ is approximate central angle between the locations in radians computed from the location-hashes and $\theta_t$ is true central angle computed from actual location values. Table~\ref{tab:location_hash_accuracy} shows that both the MAE and RMSE reduces with larger $K$ values. The error is larger than the actual distance of $25$km for $K$ smaller than 2048. In this paper, we fix $K=4096$ for our experiments in Section \ref{sec:experimentalresults}.

\subsection{Categorical Data}
\label{sec:categorical}
For categorical data types, HMAC-SHA256 is computed upon each data sample, which is keyed under a 128-bit long-lived key, $HK$, that is generated randomly upon initialisation by the ConSec mobile application. $HK$ is unique to each application and should, ideally, be generated and stored securely in the device's keystore.  The following section describes how HMAC and SB-LSH protected data is used by the ConSec authentication algorithm.


\section{Authentication Algorithm}\label{sec:contextauthenticationalgorithm}
This section details how users' authentication models are learned from SB-LSH and HMAC-applied contextual data.
\subsection{Feature Extraction}
\label{sec:feature_extraction}
\begin{table}
	\centering
	\caption{ConSec contextual data modalities.}
	\label{tab:contextual_feature_types_location}
	\begin{threeparttable}
		\begin{tabular}[t]{ p{3.2cm} c c}
			\toprule
			\textbf{Contextual Modality} & \textbf{Feature Type} & \textbf{Data Type}\\ \midrule
			Location-hash & $\dag$ & N \\ 
			Wi-Fi state & OHE & C \\
			Wi-Fi router MAC & OHE & C \\
			Wi-Fi SSID & OHE & C \\
			Wi-Fi IP & OHE & C \\ 
			Wi-Fi network ID & OHE & C \\
			Wi-Fi RSSI  & V & N \\
			Wi-Fi frequency & OHE & C \\
			Wi-Fi router IP & OHE & C \\ 
			Wi-Fi router DNS-1 & OHE & C \\ 
			Wi-Fi router DNS-2 & OHE & C \\ 
			List of Wi-Fi names & OHE & C \\ 
			List of Wi-Fi MACs & OHE  & C \\ 
			List of Wi-Fi frequencies & OHE & C \\
			SIM state & OHE & C \\ 
			Network data state & OHE & C \\ 
			Network data type & OHE & C\\ 
			Network RSSI & V & N\\ 
			Network operator name & OHE & C\\ 
			Network MCC MNC & OHE & C \\ 
			Network LAC & OHE & C\\ 
			Network Cell ID & OHE & C \\
			Bluetooth device names & OHE & C \\
			Bluetooth device MACs & OHE & C\\ 
			Day index & OHE & C\\ 
			Time & V & N \\ \bottomrule
		\end{tabular}
		\begin{tablenotes}
			\item $\dag$: Cosine similarity with six references, V: real value,  OHE: one-hot encoded, N: numerical, C: categorical.
		\end{tablenotes}
\end{threeparttable}
\end{table}

Table~\ref{tab:contextual_feature_types_location} lists the 32 contextual data types employed by ConSec. The LSH data is pre-processed to compute the feature wherein the cosine similarity values are computed with respect to a reference value collected during the enrollment phase. The modal (most frequent) LSH value from the enrollment samples is used as this reference.
For categorical data, one-hot encoded (OHE) features are created to map categorical values to numerical representations; OHE results in a sparse matrix as an output, where each column corresponds to one possible value of the data. Real-valued data types, indicated in Table~\ref{tab:contextual_feature_types_location}, are used directly. We note that feature vectors are standardised to zero mean and unit variance once computed.  


\subsection{Learning User Authentication Models}
Computed features are subsequently inputted to a k-means clustering algorithm, which is initialised using the k-means++ method by Arthur and Vassilvitskii~\cite{Arthur_2007}. Clustering was chosen from the assumption, based from related work~\cite{hayashi_casa:_2013,conXsense,gupta2012intuitive}, that users' contextual data has a tendency to form clusters according to their mobility patterns, such as regularly frequenting places of work and home. 

We are interested in detecting the centroid of these clusters, thus k-means was used, but this necessitates some number of clusters, $k$, to be specified in advance.  As such, the algorithm is executed for different values of $k=[5, 10, 15, 20, 25]$, stopping for a value of $k$ if the rate of the clustering error falls below a convergence threshold, $\epsilon_1$. In the experiments, a threshold of $\epsilon_1 = 0.05$ (5\%) was used.  Additionally, we prune clusters with population density lower than a separate threshold, $\epsilon_2$. We fix $\epsilon_2 = 0.025$ (2.5\%) in our experiments so that clusters with densities below this, i.e. very rarely or transiently frequented, are removed from the user model.

\subsection{Authentication Procedure}
As shown in Figure~\ref{fig:contextual_authentication_blockdiagram}b, ConSec-Auth is ready to enter the authentication phase once the user models are created. This comprises an eight step process illustrated in Figure~\ref{fig:info_flow} and described as follows:

\begin{figure}
    \centering
    \includegraphics[width=\linewidth]{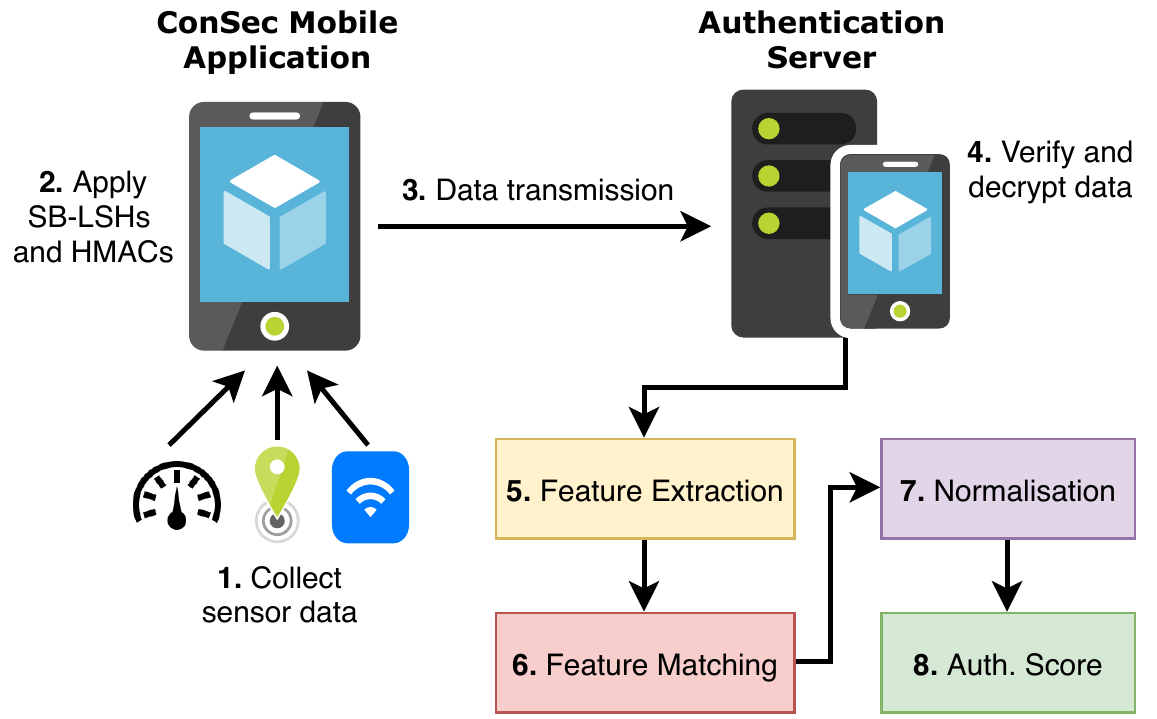}
    \caption{Information flow for user authentication.}
    \label{fig:info_flow}
\end{figure}

\begin{enumerate}
    \item \textbf{Data Collection}. Data is sampled by ConSec-App at 10 minute intervals using the modalities listed in Table~\ref{tab:contextual_feature_types_location}.
    \item \textbf{Device Pre-Processing}. Next, ConSec-App applies SB-LSHs and HMACs to numerical and contextual modalities respectively, as described in Sections~\ref{lsh:described} and~\ref{sec:categorical}.
    \item \textbf{Data Transmission}. The SB-LSH and HMAC values are transmitted over a secure channel between ConSec-App and ConSec-Auth using TLS. 
    \item \textbf{Verification and Decryption}. The server verifies and recovers the device feature vectors from the secure channel.  
    \item \textbf{Feature Extraction}. This follows the procedure described in Section~\ref{sec:feature_extraction} to produce a feature vector, $\bf{f}$.
    \item \textbf{Feature Matching}.  ConSec-Auth computes the Euclidean (L2) distance between $\bf{f}$ and the cluster centroids of the user's model. The smallest distance ($d$) is selected.
    \item \textbf{Normalisation}. The smallest distance $d$ is normalized as $\frac{d}{q}$, where $q$ is a normalizing constant. The score is computed as $\mathcal{S} = 1 - \frac{d}{q}$. 
    \item \textbf{Score Output}. The authentication score, $\mathcal{S} \in [0,1]$, is returned as output.
\end{enumerate}

\subsection{Model Refreshing}

Some contextual data may change over time due to behavioural shifts in the users' mobility patterns; for example, after joining a new workplace or moving house in a new town/city where the old locations have little relevance.  This applies not only to changes in the users' GPS locations, but also the lists of detected Wi-Fi APs, nearby Bluetooth devices, background noise, and other modalities; users' authentication data models should be updated regularly to reflect these changes.  In this work, a weekly interval was chosen as a default to re-evaluate behavioural shifts; however, the model of the previous week is retained if 33.3\% of the contextual data is predicted with higher accuracy (assuming the user spends at least 8 out of 24 hours at a regular location, e.g. home or office).

\section{Experimental Results}\label{sec:experimentalresults}
In this section, experiments are presented for evaluating the robustness of ConSec, which were conducted using data collected from 35 participants.



\subsection{Data Collection}

Participation emails were sent to 250 employees within a European-based US technology company, resulting in 35 users who enrolled in the trial; no incentives were offered for participation.  Participants were requested to install and launch the ConSec application on their primary smartphone, after which no further interaction was required; users were able to start/stop data collection at will and withdraw at any time by uninstalling the application.   Prior to this, participants were informed about types of data collected; that it would be stored alongside an anonymous, randomly-generated user ID; and an explanation of how the collected data itself was protected on the server. Supplementary material was also distributed via email. The data collection period lasted five weeks.

Upon installation, the application randomly generated unique SB-LSH parameters and $HK$ for the keying HMACs, which were held only on the device. The application collected contextual data at 10 minute intervals, which is encrypted under AES in GCM mode using a key, $AK$, generated randomly for each message. $AK$ is then encrypted using asymmetric encryption (RSA-2048) using the ConSec certified public-key, $PK$, contained within the application, and transmitted to the server which possesses the private portion. Both the data and the encrypted $AK$ are transmitted as a single message, $m$, over TLS to the authentication server. This is listed in Equations~\ref{eq:message} and~\ref{eq:message1}. This process was conducted to protect participants' data at rest on our authentication server.
\begin{multline}
    \begin{array}{l@{\hspace{2pt}}l}
    Data = \text{AES-GCM}_{AK}\big\{
    &\mathrm{LSH}(p_x, p_y, p_z, \text{Alt.}, \text{Noise},\dots),\\
    &\mathrm{HMAC}_{HK}(\text{Wi-Fi}_{\text{MAC}}),\\
    &\mathrm{HMAC}_{HK}(\text{Wi-Fi}_{\text{ESSID}}),\dots\big\}
    \end{array}
    \label{eq:message}
\end{multline}

\begin{equation}
    m = \big\{Data,\, \text{RSA-Enc}_{PK}(AK)\big\}
    \label{eq:message1}
\end{equation}

To minimise battery consumption, the ConSec-App placed sensors into low-power mode due to the relatively low sampling frequency (once every 10 minutes).  Preliminary experiments on a single device showed that ConSec-App consumes a negligible ($\sim$1\%) amount of power, indicated by the power management metrics provided by Android OS on our test device (Samsung Galaxy S8). The encrypted contextual data was uploaded to a server on a weekly basis over TLS with an anonymous user ID generated by ConSec-App.   The encrypted data was stored on a private corporate machine, with access granted only to the authors of this work.

\subsection{Robustness to Inaccurate Inputs}
\label{sec:sensitivity}

\begin{figure}
	\centering
	\includegraphics[width=3.5in]{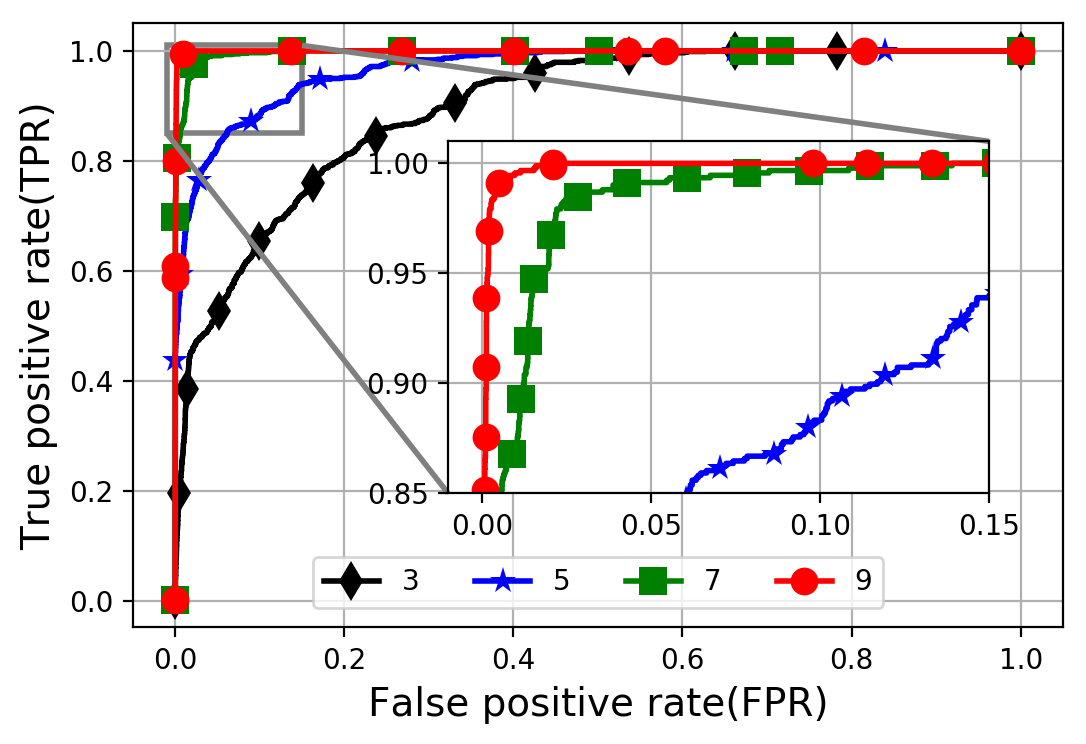}
	\caption{ROC curves for (a) 3, (b) 5, (c) 7 and (d) 9 numbers of the modified contextual data modalities (averaged across all participants).}
	\label{fig:sensitivity_to_incorrect_ctxt_fields}
\end{figure}

Unlike previous approaches that rely upon \emph{supervised} learning, e.g. CASA~\cite{hayashi_casa:_2013}, ConXsense~\cite{conXsense} and SenGuard~\cite{shi_senguard:_2011}, whereby the algorithm is trained using labels collected from user input, ConSec is underpinned by clustering, i.e. unlabelled \emph{unsupervised} learning.  Evaluating the performance of clustering algorithms is non-trivial compared with computing the precision, recall, or similar metric drawn from counting the number of true/false positives/negatives used in previous work~\cite{scikit-learn}.  

In this paper, we began our evaluation by measuring the robustness of ConSec to inaccurate values sourced from varying numbers of data modalities. This is to evaluate its robustness to values modified by a contextual adversary listed in Section~\ref{sec:threat_model}. To this end, Receiver Operating Characteristic (ROC) curves were plotted for the True Positive Rate (TPR) and False Positive Rate (FPR) to evaluate the detection of compromised samples versus unmodified ones.



For this, we selected the 100 closest feature vectors to the cluster centroids using the L2-distance from the training sets of each user. These samples are labelled as positives that should be accepted by the system. Next, $n$ feature fields of these samples are modified---using the procedure described below---and labelled as negatives that ought to be rejected by the system. This process was repeated for 1000 distinct trails for various combinations of $n=[3,5,7,9]$ modalities and across all users.


To modify the categorical data fields, such as Wi-Fi ESSIDs, the HMAC values (base 64-encoded) were modified to a random base 64 string of the same length. For numerical data, the SB-LSH values were modified to a random binary string, also of the same length. Figure~\ref{fig:sensitivity_to_incorrect_ctxt_fields} shows the ROC curves using vectors with three, five, seven and nine contextual modalities. 
Our results show that the system approaches an ideal system when seven and nine feature fields are modified; that is, the experiments show that erroneous measurements can be detected with TPR=0.975 and FPR=0.025 for seven affected modalities. This error rates increase significantly as the number of affected fields decreases; the system is less robust when $n=3$, with approximately TPR=0.8 and FPR=0.2.   

\section{Privacy and Security Analysis}\label{sec:privacyanalysis}
In this section, we present informal and experimental analyses of the privacy and security properties offered by ConSec.

\subsection{Privacy Analysis}

For categorical modalities, ConSec uses HMACs directly to model user behaviour, which are keyed under a randomly generated, application-specific key, $HK$, assumed to be stored securely on the device. Assuming the use of a secure cryptographic hash function, this avoids disclosing the actual underlying values to an observant server. Moreover, the server does not possess $HK$, thus precluding the use of dictionary attacks to which non-keyed functions would be vulnerable.

Regarding SB-LSH for numerical data, we note that some information could be leaked based on SB-LSH hashes leaked previously in conjunction with knowledge of their distances in reality. While the actual locations on Earth's surface are masked, SB-LSH reveals the {\it{angular distance}} between them. This raises the following question: assuming that some SB-LSH values taken at points during the day can be mapped to known locations---for example, the user being at work during the day and at home at midnight---can one deduce other locations using triangulation? 

\begin{figure}
	\centering
	\includegraphics[width=2in]{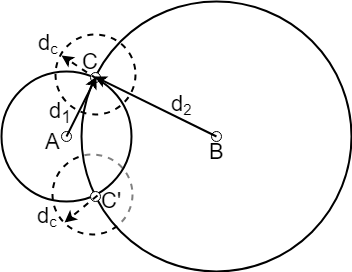}
	\caption{Triangulation to derive unknown location, $C$, from two known locations, $A$ and $B$. Only distances from SB-LSH values $d_1$ and $d_2$ are known for $C$ with respect to $A$ and $B$.}
	\label{fig:triangulation}
	
\end{figure}

Suppose locations $A$ and $B$ correspond to the user's home and office locations respectively, and the actual GPS coordinates of these locations are known. The issue pertains to the precision and accuracy with which the latitude and longitude values of unknown location $C$ can be triangulated. From the SB-LSH values of $A$, $B$ and $C$, one can compute distances $d_1$ (between $C$ and $A$) and $d_2$ (between $C$ and $B$) and solve for the latitude and longitude values for unknown location $C$. This can result in locations either $C$ or $C'$. We use the chord distance given in Equation \ref{eq:distance_from_angle} to compute the distance between the locations, where the approximate angle $\theta_a$ can be computed from the location-hashes and, $R$, the mean value of the radius of Earth. The precision and accuracy to which $C$ can be localised depends on how accurately and precise $d_1$ and $d_2$ can be computed from SB-LSH values. Ultimately, $C$ may be computed with some precision and accuracy that is determined by a confusion region with radius $d_c$. The larger error in computing $d_1$ and $d_2$, the larger $d_c$, thus decreasing the precision to which $C$ can be localised and, hence, offering a greater degree of privacy to the user.  We illustrate this in Figure~\ref{fig:triangulation} for clarity.
\begin{table}
	\centering
	\caption{Mean Absolute Error (MAE), Root-Mean-Square-Error (RMSE), Mean, and Standard Deviation of distances errors from SB-LSH values for varying actual distances in kilometers. (All values given w.r.t. each actual distance).}
	\label{tab:lsh_privacy_analysis}
	\begin{tabular}[t]{ c c c c c} \toprule
		\makecell{Actual\\distance (km)} & MAE & RMSE & Mean & Std. Dev. \\ \midrule
		5 & 2.826 & 3.266 & -1.109 & 3.804 \\ 
		10 & 3.908 & 4.776 & -1.645 & 4.616 \\
		25 & 9.785 & 12.648 & -2.643 & 12.344 \\
		50 & 14.779 & 18.819 & -4.683 & 18.588 \\
		100 & 26.258 & 33.389 & -9.417 & 29.975 \\
		500 & 100.980 & 124.377 & -51.619 & 113.846 \\
		1000 & 181.192 & 230.074 & -96.341 & 212.993 \\ \bottomrule
	\end{tabular}
	\vspace{-1cm}
\end{table}

We show that the errors in computing distances, e.g. $d_1$, from the location-hashes are non-uniform and increase for larger actual distances. To show this, we perform simulations using randomly generated pairs of latitude and longitude values with a fixed distance between them. The SB-LSH hashes are computed for the pair values, and the distance between the locations is computed from SB-LSH values. The distances are computed using Equation \ref{eq:error_in_distance}. Table \ref{tab:lsh_privacy_analysis} shows the MAE, RMSE, mean, and standard deviation of the errors in computing distances for differing (actual) distances between the pairs of $d=[5,10,25,100,500,1000]$ kilometers. 

Our results show that, as expected, the error is larger when the location is at a greater actual distance. For example, when $C$ is at 5km from $A$, the error in computing the distances from the SB-LSH values has MAE=2.826km; at an actual distance of 25km, this increases to MAE=9.785km. Thus, $C$ is localised with greater error, and with poorer accuracy, when it is located further away in reality. 

\begin{figure}[!ht]
	\centering
	\includegraphics[width=\linewidth]{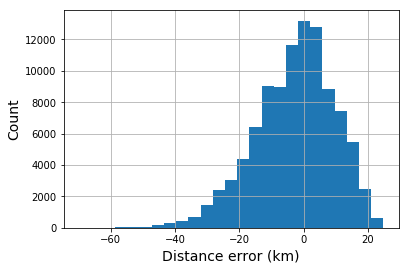}
	\caption{Histogram of errors in computing distances from SB-LSH values w.r.t. points at an actual distance of $25$~km.}
	\label{fig:lsh_privacy_analysis_distribution_25}
\end{figure}

\begin{figure}[!ht]
	\centering
	\includegraphics[width=\linewidth]{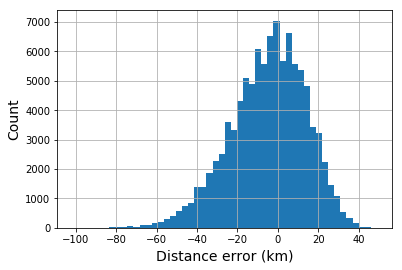}
	\caption{Histogram of errors in computing distances from SB-LSH values w.r.t. points at an actual distance of $50$~km. 
	}
	\label{fig:lsh_privacy_analysis_distribution_50}
\end{figure}

Understanding the actual distribution of the distance errors is also important; uniformly distributed error is desirable such that any unknown locations cannot be localised with high probability. Figures~\ref{fig:lsh_privacy_analysis_distribution_25} and~\ref{fig:lsh_privacy_analysis_distribution_50} show the histogram of distance errors for actual distance of $25$ and $50$ kilometers respectively using a bin size of $1000$ meters. The histogram of the distance error is approximately Gaussian, with a flat peak and wide variance; the greatest distances fall at the tails of the distributions. That is, unknown locations that are far away cannot be localised with low error. 

From the above, we can conclude that inferring locations from SB-LSH values incurs significant error, even if some information is leaked about the user with the assistance of supplementary knowledge about the users' actual locations in reality. This uncertainty is determined primarily by the largest distance of the two closest known locations. The greatest distance determines the error ($d_c$ in Figure~\ref{fig:triangulation}) in estimating the unknown locations from SB-LSH hashes. Moreover, the error increases dramatically and proportionally to larger actual distances in reality. 

\subsection{Security Analysis}
We now describe how ConSec thwarts a variety of security threats drawn from the threat model in Section \ref{sec:threat_model}.


\subsubsection{Modified Input Modalities}
One attack pertains to modifying the SB-LSH and HMAC values produced by ConSec-App with the aim of poisoning the model to deny future legitimate accesses or facilitate illegitimate accesses.  While data is secured within transit using TLS, as described in Section \ref{sec:contextauthenticationalgorithm}, this does not preclude an adversary with the ability to alter some number of SB-LSH and HMAC values within ConSec-App itself. Our experimental results in Section \ref{sec:sensitivity} show that seven or more inaccurate modalities (of 26 in total) can be detected with low error (approximately TPR=0.975 and FPR=0.025), which somewhat decreases when a smaller number of inputs are modified (TPR=0.8, FPR=0.2 for three inputs).

\subsubsection{Server-side Privacy}
Evidently, using raw contextual values gives rise to privacy risks against honest-but-curious adversaries who aim to exploit users' behavioural data for ulterior purposes. This extends to unauthorised disclosure if an attacker successfully exfiltrates data from the server and releases it publicly after exploiting some vulnerability.  We reduce the impact of these cases as ConSec models and authenticates users' behaviour \emph{directly} from keyed HMAC and SB-LSH values rather than raw, unprotected values. In Section \ref{sec:experimentalresults}, we also show experimentally that significant errors are involved when attempting to infer unknown locations from known SB-LSH values.

\subsubsection{Network Attacks}   
Another threat to contextual authentication systems is where a network adversary re-sends previously observed messages containing the feature vectors (whether protected or not) to the server, with the aim of poisoning the behavioural model and influencing the authentication algorithm. This attack is protected through the use of a secure channel (TLS) between the application and authentication server with replay protection.  Moreover, the contextual data vector includes a timestamp to ensure its recentness against replay attacks.

\section{Conclusion}\label{sec:conclusion}
In this work, we presented ConSec---a privacy-enhancing, context-aware authentication system that utilises locality-sensitive hashing for masking plain sensor measurements from a user device. ConSec learns, models and authenticates users' behaviour directly from protected values in an outsourced fashion without disclosing raw measurements to an honest-but-curious authentication server.  We began with a detailed review of existing contextual authentication schemes and methods for preserving the privacy of sensitive input modalities for continuous authentication generally.   We then showed how ConSec supports learning from both numerical and categorical data in a flexible manner using the SB-LSH algorithm by Ji et al.~\cite{ji_super-bit_2012} and keyed HMACs, without incurring the complexities of existing privacy-enhancing approaches, such as homomorphic encryption and multi-party computation.  After describing the data types and authentication algorithms used by ConSec, experimental results were presented using data collected from 35 users in the field.  This was followed by analyses showing its its robustness to modified input modalities and the errors involved in attempting to infer unknown locations from known values.  

In future work, we aim to evaluate ConSec in a long-term usability study to determine users' attitudes towards using the system and, in particular, concretely measuring its handling of gradual behavioural shifts in users' behavioural patterns. We also aim to explore the scheme's robustness to \emph{mimicry attacks} in which a physical attacker possesses the device and attempts to reconstruct the victim's behaviour, such as visiting their home or place of work. 

\begin{acks}
The authors would like to thank other members of the OneSpan Innovation Centre---Paul Dunphy, Andreas Gutmann, Sharon Lee, Steven J. Murdoch, and  Tom De Wasch---and the reviewers for providing many useful comments and suggestions towards improving this work. 
\end{acks}
\balance
%
\bibliographystyle{ACM-Reference-Format}
\bibliography{context_bib}

%


\end{document}